\documentclass[10pt]{iopart}

\usepackage{graphicx}
\usepackage[usenames]{color}
\usepackage{amssymb}

\begin{document}

\title{Stable controllable giant vortex  in a trapped Bose-Einstein condensate}

\author{ S. K. Adhikari\footnote{Accepted in Laser Physics Letters}\footnote{URL: professores.ift.unesp.br/sk.hadhikari}
} 
 
\address{
Instituto de F\'{\i}sica Te\'orica, UNESP - Universidade Estadual Paulista, 01.140-070 S\~ao Paulo, S\~ao Paulo, Brazil
} 
\ead{sk.adhikari@unesp.br}

\begin{abstract}
 
In a harmonically-trapped rotating Bose-Einstein condensate  (BEC),  a vortex of large angular momentum decays to multiple vortices of unit angular momentum from an energetic consideration.  
We demonstrate the formation of a robust and dynamically stable    giant vortex of large angular momentum  in a harmonically trapped rotating 
BEC with a potential hill at the center, thus forming a Mexican hat like trapping  potential.
For a small inter-atomic interaction strength,  a  highly controllable stable giant vortex appears,  whose angular momentum slowly increases as the  angular frequency of rotation is increased.   As the   inter-atomic interaction strength is increased beyond a critical value, only  vortices of unit angular momentum  are formed, unless the strength of the potential hill at the center is also increased: for a stronger potential hill at the center a  giant vortex is again formed. The dynamical stability of the giant vortex is demonstrated by real-time propagation numerically. 
These giant vortices of large angular momentum  can be   observed and studied experimentally   in a highly controlled fashion.

\end{abstract}



\maketitle

\section{Introduction}

 Soon after the observation  \cite{becexpt}
 of     trapped Bose-Einstein condensates (BEC) of alkali-metal atoms \cite{yuka} in a laboratory, rapidly rotating 
trapped condensates  were created and studied. A small number of vortices were created \cite{vors} for a small angular frequency of rotation $\Omega$. 
As the angular frequency of rotation is increased in the rotating BEC,   energetic consideration favors the  formation of  a lattice of quantum vortices of unit angular momentum  each ($l=1$) per atom \cite{abri,fetter} and not an angular momentum state with $l>1$. This was first confirmed experimentally  in liquid He II in bulk \cite{4he} and later in { a} dilute trapped BEC \cite{vors,vorl}.
Consequently, a rapidly rotating trapped BEC generates  a large number of vortices of unit angular momentum  { usually arranged} in a Abrikosov  triangular lattice \cite{abri,vorl}.   The dilute trapped BEC is formed  in the perturbative weak-coupling   mean-field limit. This allows      to study the formation of  vortices in such a   BEC  by the mean-field   Gross-Pitaevskii (GP) equation \cite{GP}.

There has also been a study of vortex-lattice formation in a BEC along the weak-coupling to unitarity crossover \cite{sci}.
 The study of vortex lattices in a binary or a multi-component spinor BEC is also interesting because the interplay between intra-species and inter-species interactions may lead to the formation of  square    \cite{Schweikhard,kumar}, stripe  and honeycomb 
 \cite{honey} vortex lattice, other than the standard Abrikosov triangular lattice \cite{abri}.    In addition, there could be
the formation of coreless vortices \cite{coreless}, 
  vortices of fractional angular momentum \cite{frac1},   and phase-separated vortex lattices in multi-component non-spinor \cite{cns},
spinor  \cite{thlatsep} and dipolar  \cite{kumar}   BECs.

The  challenging task of the formation of a   giant vortex with a large angular momentum in a trapped BEC is of interest to both theoreticians and experimentalists. Such a vortex could be of use in    quantum information processing
technology.  Dynamically formed 
meta-stable giant vortices of angular momentum  7 to 60   were experimentally observed \cite{cornell} and studied  numerically \cite{simula}.  There have been
numerical studies of a giant vortex in a {  quartic plus quadratic (harmonic)  \cite{engels,aftalion}  and also quartic minus quadratic  \cite{aftalion,kasamatsu} traps.  Such an asymptotically quartic trap is difficult to realize in a laboratory. 
 In most of } these studies 
the giant vortex appeared as a hole in a BEC  with large inter-atomic interaction strength,   surrounded by a vortex lattice with a large number of vortices.  There { have also} been studies of a giant vortex in multi-component BECs \cite{multi}. 
In { most of} these studies there was no control over the total number of vortices and the total angular momentum associated with the giant vortex. Such a vortex state cannot be employed in precision studies.

In the present letter we suggest a way of generating a stable giant vortex in a BEC with very small inter-atomic interaction strength, rotating with a small angular frequency of rotation $\Omega$. The trapping potential was essentially harmonic with a small hill at the center in the shape of a Mexican hat. Such a potential can be optically 
realized in a laboratory { by a blue-detuned, Gaussian laser beam} \cite{detuned}.
 With the increase of $\Omega$, the angular momentum in the  giant vortex can be increased gradually in a controlled fashion and can be fixed at any desired value.  This control over the angular momentum  of giant vortices and the associated small inter-atomic interaction  strength {make these giant vortices}    { appropriate}  for high precision studies.

In section 2  the mean-field model for a   rapidly rotating binary BEC is presented.  Under a tight trap in the transverse direction a quasi-two-dimensional (quasi-2D) version of the model is also given, which we use in this letter. 
The results of numerical calculation are shown in section 3.  
Finally, in section 4 we present a brief summary of our findings.

\section{Mean-field model for a rapidly rotating binary BEC}

The generation of quantized vortices in a dilute BEC or in liquid He II upon rotation is an earmark of super-fluidity.  As suggested by Onsager \cite{onsager}, Feynman \cite{feynman} and   Abrikosov \cite{abri},   
 the vortices in a rotating super-fluid 
have quantized circulation: \cite{fetter,Sonin}
\begin{equation}\label{qv}
\oint_{\cal C} {\bf v} .  d{\bf r}=\frac{2\pi\hbar l}{m},
\end{equation}
where  ${\bf v}({\bf r},t)$ is the super-fluid velocity at space point ${\bf r}\equiv (x,y,z)$ and {at}
time $t$,  ${\cal C}$ is a generic closed path, 
  $l$ is  the quantized integral  angular momentum of an atom in units of $\hbar$  and $m$ is the mass of an atom. 
 In a rotating super-fluid, the integral (\ref{qv}) over 
path ${\cal C}$ could {be}
nonzero,  implying a  topological defect in the form of quantized vortex inside this path.   The quantization of circulation was explained 
by London   assuming that the super-fluid dynamics  is driven by the  complex scalar field \cite{fetter,phase}
\begin{equation}
\phi({\bf r},t)= 
 |\phi({\bf r},t)|   e^{i \varphi({\bf r},t) },
\end{equation}
 which satisfies the nonlinear mean-field GP equation \cite{fetter}
\begin{eqnarray}\label{gpeq}
{\mbox i} \hbar \frac{\partial \phi({\bf r},t)}{\partial t}  &=&
{\Big [}  -\frac{\hbar^2}{2m}\nabla^2
+ \frac{1}{2}m \omega^2 \kappa^2
{z}^2+V(x,y ) \nonumber \\ &+&  \frac{4\pi \hbar^2}{m}{a} N \vert \phi({\bf r},t)\vert^2
{\Big ]}  \phi({\bf r},t),
\end{eqnarray}
{ where   $\varphi$ is the phase of the function $\phi$, 
$N$} is the number of atoms, $\kappa \omega$ is the harmonic trap frequency in the $z$ direction with the constant $\kappa\gg 1$, $a$ is the atomic scattering length and  $V({x,y})$ is the Mexican hat  trapping potential in the $x-y$ plane:
\begin{eqnarray}
V(x,y)=\frac{1}{2}m\omega^2(x^2 +y^2)+A\exp^{-(x^2+y^2)/4},
\end{eqnarray}
where the central hill of height $A$ is taken to be a Gaussian.
Such a Gaussian potential can be realized in a laboratory by de-tuned   laser beams \cite{detuned}.
  The function $\phi$ is normalized as $\int d{\bf r}|\phi({\bf r},t)|^2 =1.$

  In the rotating frame  of reference  the generated vortex  
state is a stationary state \cite{fetter}, which   can be obtained numerically  by
solving the underlying mean-field GP equation of the trapped BEC in the rotating frame by  imaginary-time propagation \cite{imag}.   The Hamiltonian in the rotating frame is given by \cite{ll1960} $H = H_0-\Omega l_z$, where $H_0$ is the same  in the laboratory frame, $l_z \equiv {\mbox i}\hbar (y\partial/\partial  x - x \partial/\partial y )$    the $z$ component of angular momentum.
The above transformation   suggests that the ground-state
energy 
of a  BEC  in the rotating frame should decrease  as $\Omega$ is increased {\cite{fetter,feynman}}
 and this will be verified in our numerical {calculation}.     
Using this transformation,    the mean-field 
GP equation (\ref{gpeq}) for the trapped BEC in the rotating frame for $\Omega <\omega $  can be written as \cite{fetter}
\begin{eqnarray} \label{eq1x} 
{\mbox i} \hbar \frac{\partial \phi({\bf r},t)}{\partial t}  &=& 
{\Big [}  -\frac{\hbar^2}{2m}\nabla^2 -\Omega l_z 
+ \frac{1}{2}m \kappa^2\omega^2 
{z}^2+V(x,y) \nonumber \\ &+&  \frac{4\pi \hbar^2}{m}{a} N \vert \phi({\bf r},t)\vert^2
{\Big ]}  \phi({\bf r},t).
\end{eqnarray}
However,  we recall that  for $\Omega>\omega$  a harmonically trapped  rotating BEC makes a quantum phase transition to a non-super-fluid state, where a mean-field description of the rotating BEC might not be valid \cite{fetter,engels}.

Using 
the  transformations: ${\bf r}' = {\bf r}/a_{\mathrm{ho}}, a_{\mathrm{ho}}\equiv \sqrt{\hbar/m\omega}$, $t'=t\omega,   \phi'=   \phi a_{\mathrm{ho}}^{3/2},  
\Omega'=\Omega/\omega, $ etc., a dimensionless form of  (\ref{eq1x})  can be obtained:   
\begin{eqnarray}& \,
{\mbox i} \frac{\partial \phi({\bf r},t)}{\partial t}=
{\Big [}  -\frac{\nabla^2}{2 }
+V(x,y)+ \frac{1 }{2}\kappa^2  z^2  -\Omega  l_z 
+ 4\pi Na \vert \phi \vert^2 
{\Big ]}  \phi({\bf r},t),
\label{eq3} 
\end{eqnarray}  
where  we have omitted the prime from the transformed variables.

\begin{figure}[!t]

\begin{center}
\includegraphics[trim = 0cm 0cm 0cm 0mm, clip,width=.5\linewidth]{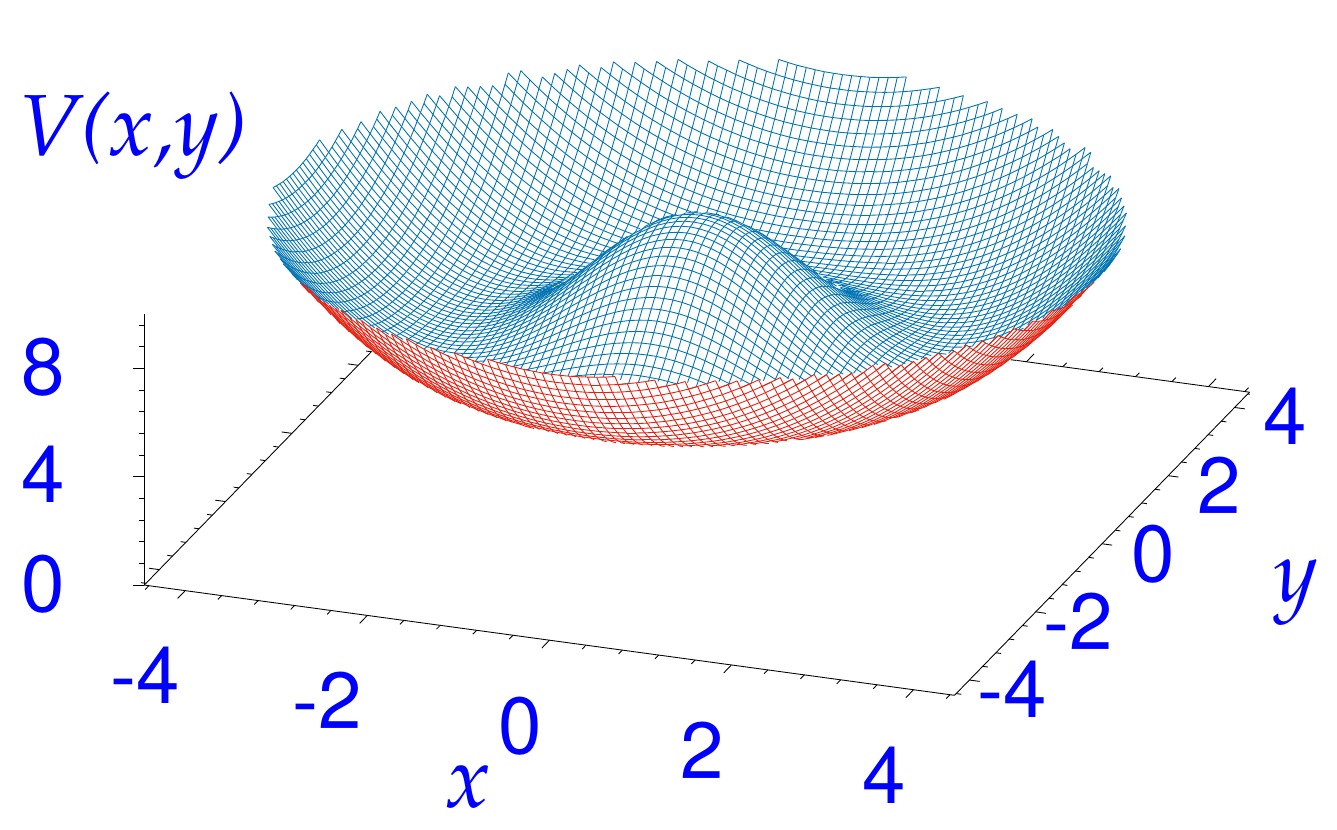} 

\caption{  Mexican hat potential $V(x,y)$ of (\ref{eq5})  with $B=10$.
}
\label{fig1}
\end{center}

\end{figure}

For a quasi-2D binary BEC in the $x-y$ plane and for $\kappa\gg 1$ the    wave function  can   be written as 
$\phi({\bf r},t)= \psi({x,y};t)\Phi(z)$, where the function $ \psi({x,y};t)$ carries the essential   dynamics and  the normalizable Gaussian function   $\Phi(z)= \exp(-z^2/2d_z^2)/(\pi d_z^2)^{1/4}, d_z= \sqrt{1/\kappa}$ plays a passive role. In this case the 
$z$ dependence can be integrated out \cite{luca} leading to  the following quasi-2D equation
\begin{eqnarray}\label{eq5x} 
{\mbox i} \frac{\partial \psi({x,y};t)}{\partial t} &=&
\left[  -\frac{\nabla^2}{2 }
+V(x,y) -{\mbox i}\Omega \Big ( \frac{y\partial}{\partial x}-   \frac{x\partial}{\partial y}  \Big ) 
+ g \vert \psi \vert^2  
\right]  \psi({x,y};t),\nonumber \\ \\
V(x,y)&=& \frac{1}{2}(x^2+y^2)+ B e^{-(x^2+y^2)/4},
\label{eq5} 
\end{eqnarray}
where  $B=A/\hbar \omega$, 
$g=2\sqrt{2\pi} a N/d_z$, and normalization $\int | \psi({x,y})|^2 dx dy =1$.    In this letter we will    consider $\Omega<1$ \cite{fetter}.   A plot of the Mexican hat potential (\ref{eq5}) for $B=10$ is shown in 
figure \ref{fig1}.

 The wave-function  $\psi(x,y,t)$ is intrinsically complex.
For a stationary state $\psi(x,y,t)\sim  \psi(x,y) e^{-{\mbox i}\mu t}$, where $\mu$ is the chemical potential.
 To write a real expression for the energy from the complex wave function $\psi(x,y)$,
it is convenient to   write   two coupled non-linear equations for the real and imaginary parts of the wave function \cite{cpckk}
\begin{equation}\label{ph}
\psi=\psi_R+\mathrm i \psi_I\equiv \sqrt{\psi_R^2+\psi_I^2}\exp(i\varphi),
\end{equation}
 where $\varphi$ is the phase of the wave function. The equation satisfied by the real part is 
 \begin{eqnarray}\label{real}
\mu \psi_R(x,y)&=& \left[-\frac{1}{2}\nabla^2+V(x,y) +g|\psi({x,y};t)|^2\right] \psi_R({x,y};t) \nonumber \\
&+&\Omega\left( y\frac{\partial} {\partial x} -x \frac{\partial}{\partial y} \right) \psi_I({x,y};t)\, .
\end{eqnarray}
In this equation $\psi_R$ is not normalized to unity.
Using ~(\ref{real}), the energy per atom for a stationary state in the rotating frame  can be expressed  as \cite{cpckk}
\begin{eqnarray}\label{e2dr}
E(\Omega)&=& \frac{1}{\int dx dy  \psi_R^2 } \int dx dy   \biggr[{\frac{1}{2}{(\nabla \psi_R)^2}}+ V(x,y)\psi_R^2 \nonumber \\ &+&
 \frac{1}{2} g(\psi_R^2+\psi_I^2){\psi_R^2} 
+\Omega    \psi_R \left( y\frac{\partial }{\partial x} -
x \frac{\partial}{\partial y} \right)\psi_I\biggr] , 
\end{eqnarray}
Equation~(\ref{e2dr})  involves algebra of real functions only and will be used in  numerical calculation.

\section{Numerical Results}

The quasi-2D  mean-field equation
 (\ref{eq5x})  cannot be solved analytically and we employ  the split time-step Crank-Nicolson method \cite{imag,CPC}  for its numerical  solution  using a space step of 0.05
and a time step of 0.0002 for imaginary-time simulation and 0.0001 for real-time simulation. The imaginary-time propagation is used to obtain the lowest-energy stationary ground state and the real-time propagation is used to study the dynamics.  There are different
C and FORTRAN programs for solving the GP equation \cite{imag,CPC}  and one should use the appropriate one.
The programs  of  \cite{imag} have recently been adapted to simulate the statics and dynamics of a  rotating BEC \cite{cpckk} and we use these {in this letter. 
 Without} considering a specific atom, we will present the results in dimensionless units for different sets of  parameters: 
$\Omega, g$. 
In the phenomenology of a specific atom, the parameter   $g$ can be varied experimentally through a variation of the underlying atomic scattering 
length 
 by the Feshbach resonance technique \cite{fesh}.

\begin{figure}[!t]

\begin{center}
\includegraphics[trim = 0cm 0.0cm 0cm 0mm, clip,width=\linewidth]{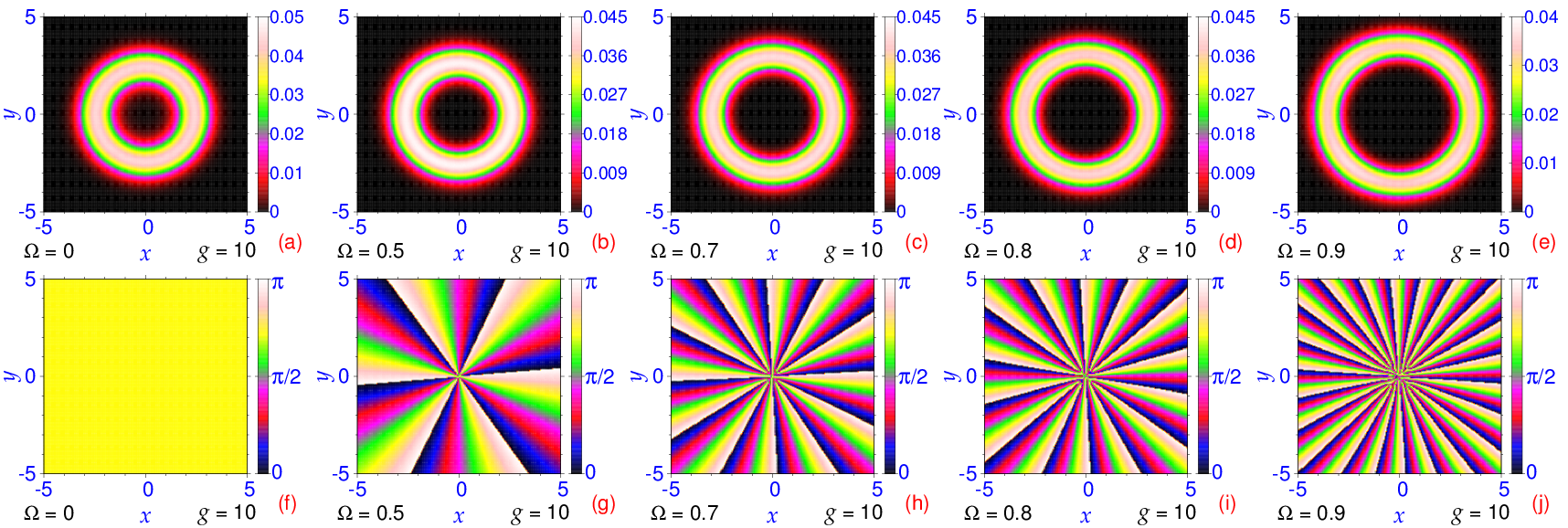}

\caption{  Giant vortex  in a  rotating quasi-2D BEC, trapped by the potential (\ref{eq5})  with $B=10$,
with $g = 10$,
satisfying (\ref{eq5x}),   from a contour
plot of 2D density $(|\psi|^2)$ for $\omega$ = (a) 0, (b) 0.5, (c) 0.7,  (d) 0.8 and (e) 0.9.
The plots in  (f)-(j) display the
corresponding phase profiles $\varphi=\arctan(\psi_I/\psi_R)$ of the rotating BECs illustrated
in (a)-(e), respectively.  All quantities plotted in this and following figures are
dimensionless.
}
\label{fig2}
\end{center}

\end{figure}

\begin{figure}[!t]

\begin{center}
\includegraphics[trim = 0cm 0.0cm 0cm 0mm, clip,width=\linewidth]{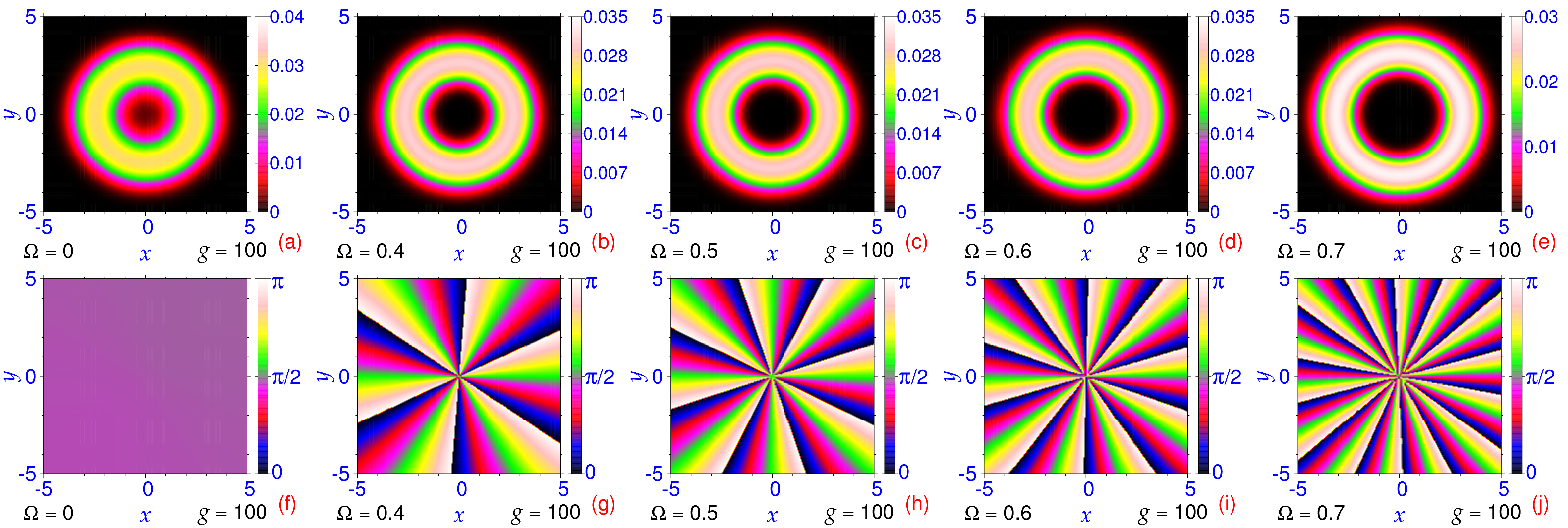}

\caption{ Giant vortex  in a  rotating quasi-2D BEC, trapped by the potential (\ref{eq5})  with $B=10$, with $g = 100$,
satisfying (\ref{eq5x}),   from a contour
plot of 2D density $(|\psi|^2)$ for $\Omega$ = (a) 0, (b) 0.4, (c) 0.5,  (d) 0.7 and (e) 0.9.
The plots in  (f)-(j) display the
corresponding phase profiles   of the rotating BECs illustrated
in (a)-(e), respectively.  
}
\label{fig3}
\end{center}

\end{figure}

We study how a giant vortex in a BEC trapped by the Mexican hat potential (\ref{eq5}) with $B=10$
 evolve with the increase of angular frequency of rotation $\Omega$.  
First we consider a small value of atomic interaction strength  $g$ $(=10)$ and solve (\ref{eq5x}) increasing  $\Omega$ gradually from 0 to 0.95.  In all cases a clean giant vortex is generated with the angular momentum increasing gradually from 0 to 11 as $\Omega$ is increased.  By the term ``clean giant vortex" we mean that there is no 
associated vortex of unit angular momentum in  the body of the giant vortex. 
The   angular momentum in the giant vortex  is obtained from an analysis of phase of the wave function. 
In Figs. \ref{fig2}(a)-(j) we plot the  density ($|\psi|^2$) of the  rotating BEC for    different $\Omega$
 and the associated phase $(\varphi)$ of the wave function, viz.  (\ref{ph}). A jump of phase of $2\pi$ along a closed path $\cal C$ around the center correspond to a unit  angular momentum, viz. (\ref{qv}).  The angular momenta of the generated giant vortex for $\omega = 0, 0.5, 0.7, 0.8$ and 0.9 are $l=$ 0, 3, 6, 7 and  11, respectively. The phases of figure 
\ref{fig2} confirm that we really have a single giant vortex  of large angular momentum and not several vortices with a large total angular momentum: this is because the phase jump takes place around a single isolated central point. 
By increasing  $\Omega$ in small steps we can easily generate 
a giant vortex with an intermediate value for angular momentum, e.g., $l=$ 2, 4, 5,  8, 9, 
10 etc. not shown in figure \ref{fig2}.

\begin{figure}[!t]

\begin{center}
\includegraphics[trim = 0cm 0cm 0cm 0mm, clip,width=.45\linewidth]{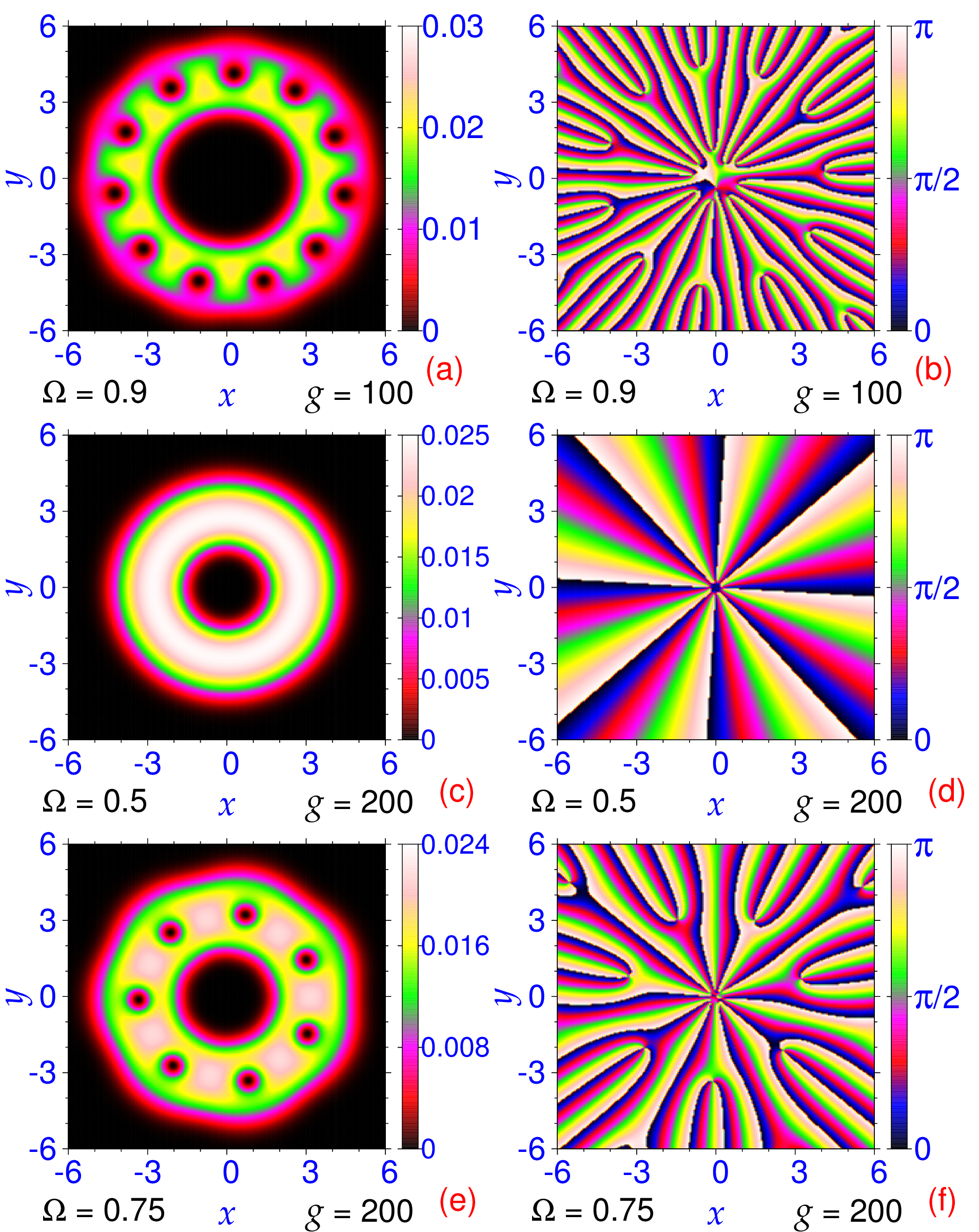} 

\caption{  Giant vortex  in a  rotating quasi-2D BEC, trapped by the potential (\ref{eq5})  with $B=10$ 
satisfying (\ref{eq5x}),   from a contour
plot of 2D density $(|\psi|^2)$ for  (a) $\Omega=0.9, g=100$, (c) $\Omega=0.5, g=200$, 
and (e)  $\Omega=0.75, g=200$. 
The plots in  (b), (d) and (f)  display the
corresponding phase profiles   of the rotating BECs illustrated
in (a), (c) and (e), respectively.  
}
\label{fig4}
\end{center}

\end{figure}

We now consider the effect of increasing the atomic interaction strength $g$. For this purpose we consider the evolution of the giant vortex for $g=100$. The numerically obtained density and the related phase profile in this case are illustrated in figure \ref{fig3} for $\Omega =0$,  0.4, 0.5, 0.6 and 0.7 in plots (a)-(e) and (f)-(j), respectively. In this case, for values of $\Omega$ up to $\Omega = 0.7$, a clean giant vortex with a large angular momentum  is obtained.  The  angular momentum  of the giant vortex for  $\Omega =0,  0.4, 0.5, 0.6$ and 0.7 are 0, 3, 
4, 5  and 7, respectively, as illustrated in figure \ref{fig3}.  It is interesting to consider the fate of the giant vortex for larger interaction strength $g$ and $\Omega$. 
 For $g=100$ and for  larger $\Omega  (\gtrapprox 0.75)$ vortices of unit angular momentum  are also generated inside the body of the giant vortex  with a large  angular momentum. This is {illustrated   in figure \ref{fig4}.}  The density and phase profile shown in figures \ref{fig4}(a) and (b) for $\Omega=0.9, g=100$   are qualitatively different from those obtained for smaller $\Omega$ and $g$. In this case we have 11 vortices of unit angular momentum  embedded in the body of the giant vortex of  angular momentum of  10 units, corresponding to a total angular momentum of 21 units.  We find that a clean giant vortex can be generated 
in a trapped BEC with $g=10$  for $\Omega \lessapprox 0.95$; for $g=100$ the same can be generated for 
  $\Omega \lessapprox 0.75$.  For $g=200$ we find that the same can be generated for   $\Omega \lessapprox 0.55$ only. 
This is illustrated in figures \ref{fig4}(c)-(f) by plots of density and phase for 
{$g=200$ and} $\Omega =0.5$ and 0.75. 
We find, from figures \ref{fig4}(c)-(d),
 that, for $\Omega =0.5, g=200$, a clean giant vortex {is} generated    with an angular momentum of 4 units, while, 
from figures \ref{fig4}(e)-(f), we find that,  for $\Omega=0.75, g=200$,  6 unit vortices are embedded in a giant vortex of angular momentum  5. For $g> 350$ no giant vortex can be generated for any $\Omega <1$ for  trapping potential 
$(\ref{eq5})$  with $B=10$.

 \begin{figure}[!t]

\begin{center}
\includegraphics[trim = 0cm 0cm 0cm 0mm, clip,width=.6\linewidth]{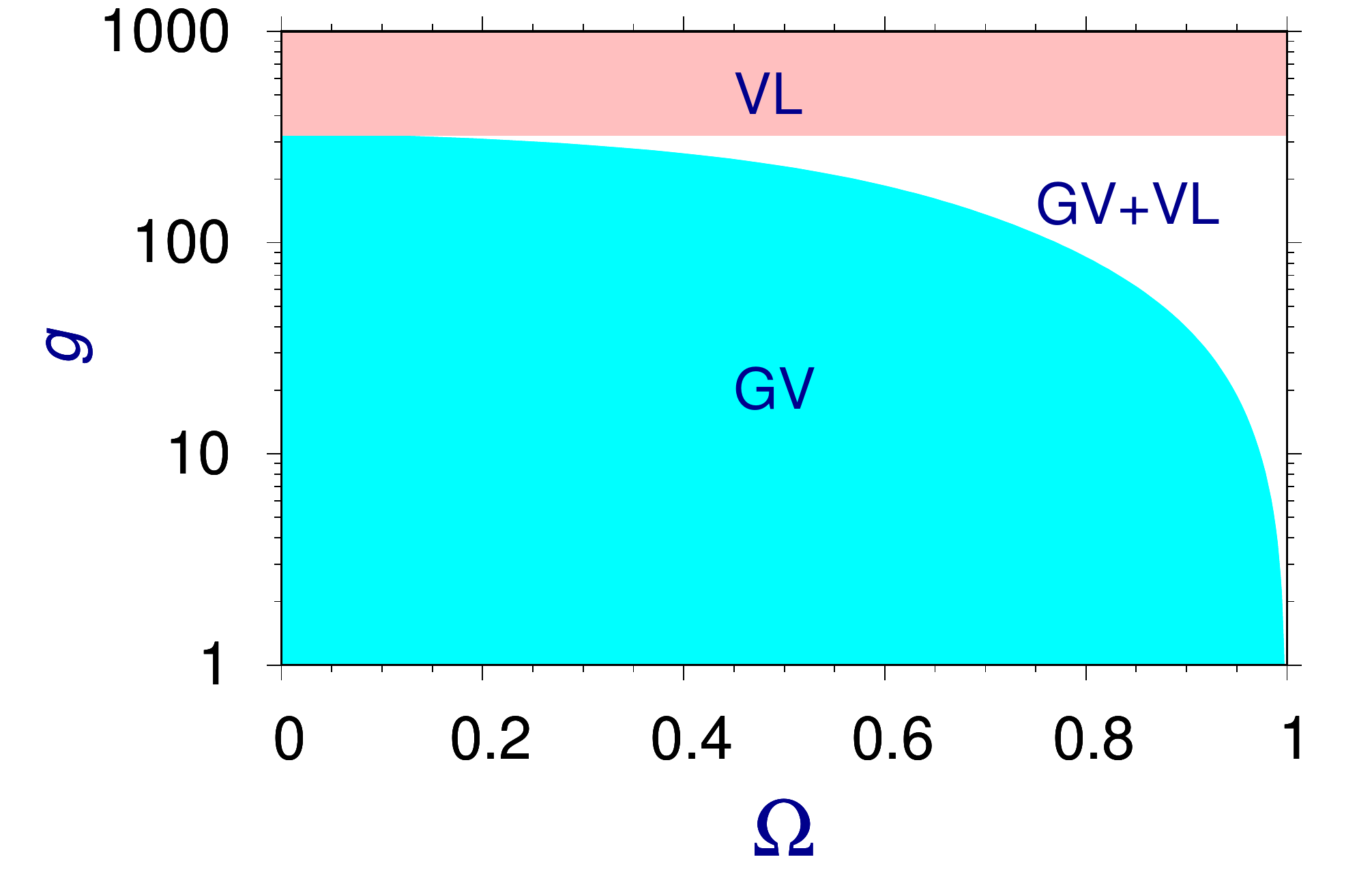} 

\caption{ A $g-\Omega$ phase plot showing the domain of formation  of clean  giant vortex (GV), giant vortex with embedded vortex lattice (GV+VL) and only vortex lattice with  no giant vortex (VL)  for the trapping potential (\ref{eq5})  with $B=10$.  
}
\label{fig5}
\end{center}

\end{figure}

\begin{figure}[!t]

\begin{center}
\includegraphics[trim = 0cm 0.cm 0cm 0mm, clip,width=.6\linewidth]{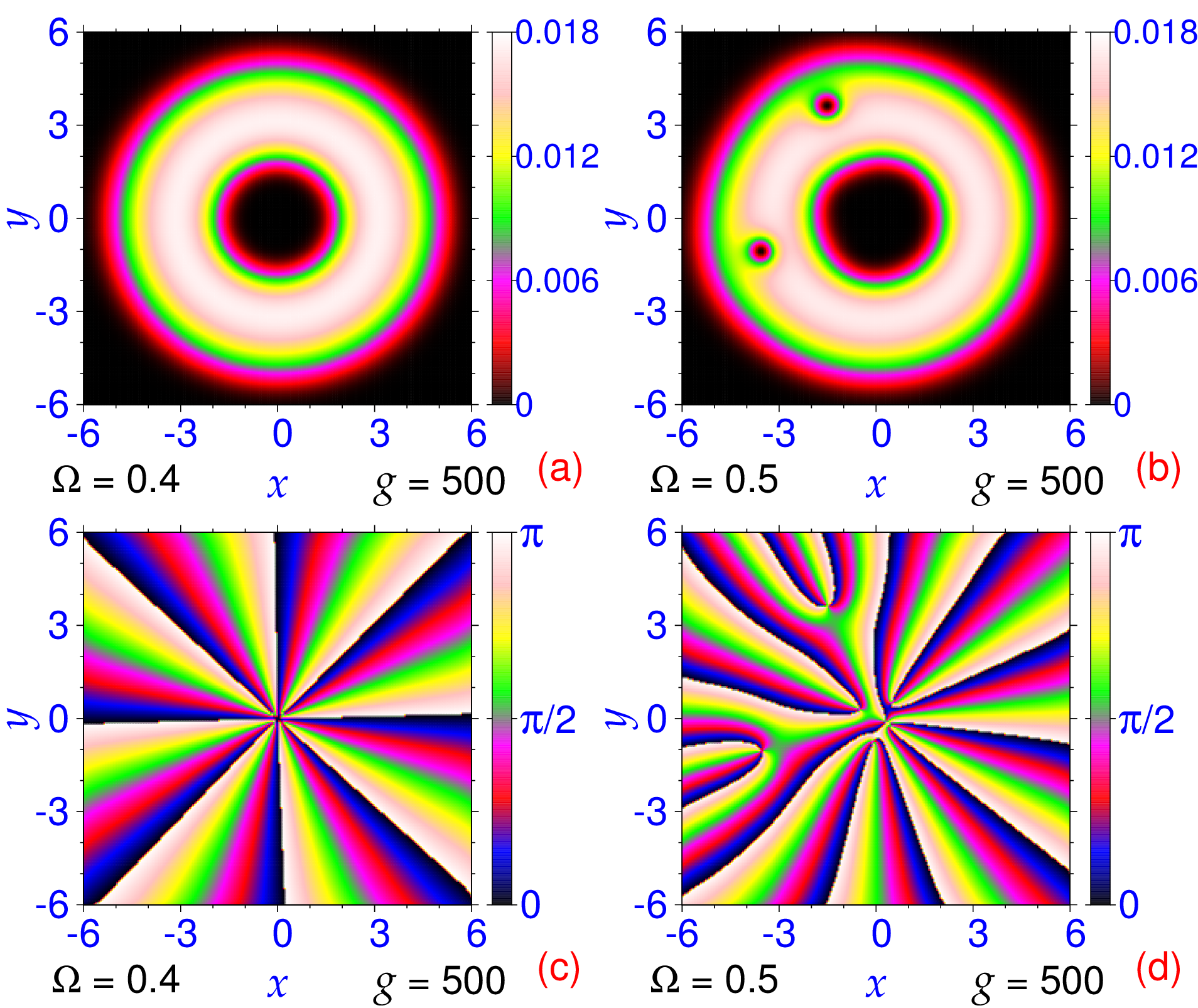}

\caption{    Giant vortex  in a  rotating quasi-2D BEC, trapped by the potential (\ref{eq5})  with $B=20$ 
satisfying (\ref{eq5x}),   from a contour
plot of 2D density $(|\psi|^2)$ for  (a) $\Omega=0.4, g=500$ and  (b) $\Omega=0.5, g=500$. 
The plots in  (c), (d)   display the
corresponding phase profiles   of the rotating BECs illustrated
in (a), (b), respectively.  
}
\label{fig6}
\end{center}

\end{figure}

\begin{figure}[!t]

\begin{center}
\includegraphics[trim = 0cm 0.cm 0cm 0mm, clip,width=.6\linewidth]{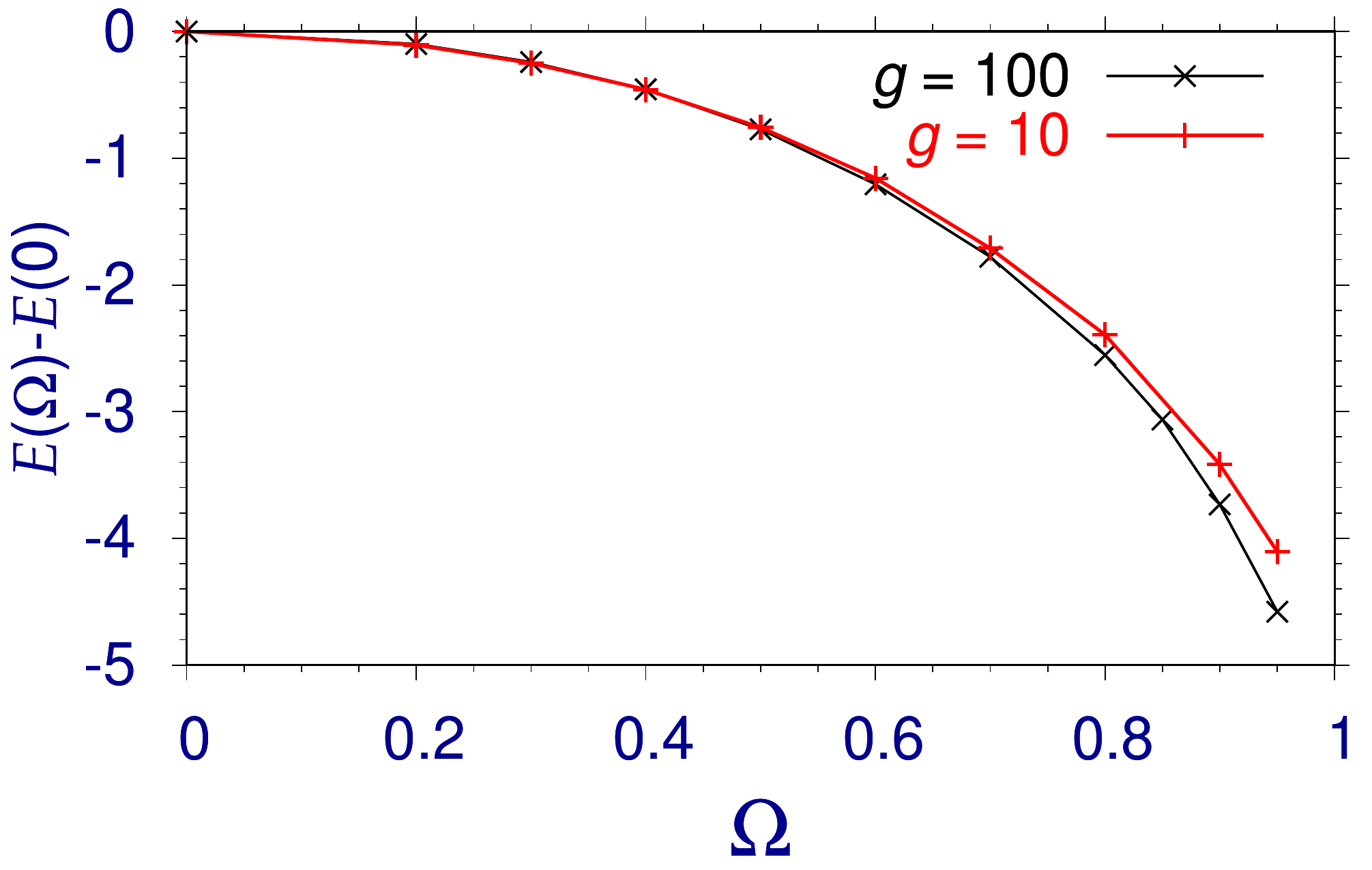}

\caption{  The rotational energy in the rotating frame $E(\Omega)-E(0)$ versus $\Omega$
for interaction strength $g=10$ and 100. The parameter $B$ of the trapping potential (\ref{eq5})
is taken as $B=10$ in both cases. 
}
\label{fig7}
\end{center}

\end{figure}

   A phase plot in the parameter space $g-\Omega$,  illustrating 
the domains   of the  formation of a clean giant vortex, a giant vortex with embedded vortex lattice, and only vortex lattice with no associated giant vortex, is exhibited in figure \ref{fig5}. From this figure we find that, for  trapping potential (\ref{eq5}) with $B=10$, it is not possible to have a giant vortex for $g>350$. However, it is possible to have a giant vortex for  a larger $g$ provided we increase the height of the hill $B$ of potential (\ref{eq5}).  For example, for $B=20$, it is possible to generate a clean giant vortex for $g=500$ as illustrated in figure \ref{fig6}, where we {show} the density and phase of the generated giant vortex for $g=500, B=20$ and $\Omega = 0.4 $ and 0.5. For $\Omega=0.4$, we have a clean giant vortex with angular momentum of 4 units as we see from plots (a) and (c). For   $\Omega=0.5$, we have two vortices of unit angular momentum embedded in the body of a giant vortex of angular momentum  of 5 units.  
Hence the domain of the giant vortex formation in the $g-\Omega$ phase plot of 
figure \ref{fig5} can be augmented by increasing the parameter $B$ in potential (\ref{eq5}).

The $\Omega$-dependent part of energy per atom can be obtained from a theoretical estimate of Fetter \cite{fetter}:
\begin{equation}\label{fet}
E(\Omega)-E(0) \approx - \frac{1}{2}I \Omega^2,
\end{equation} 
where $I$ is the moment of inertia of rigid-body rotation of an atom of the super-
fluid. We have plotted in figure \ref{fig7} this energy for $g=10 $ and 100 with the trap of 
figure \ref{fig1} with $B=10$.  The energy (\ref{fet}) is independent of $g$ and determined only by the moment
of inertia. Figure  \ref{fig7}  confirms this  universal behavior of the rotational energy.

\begin{figure}[!t]

\begin{center}
\includegraphics[trim = 0cm 0.cm 0cm 0mm, clip,width=.6\linewidth]{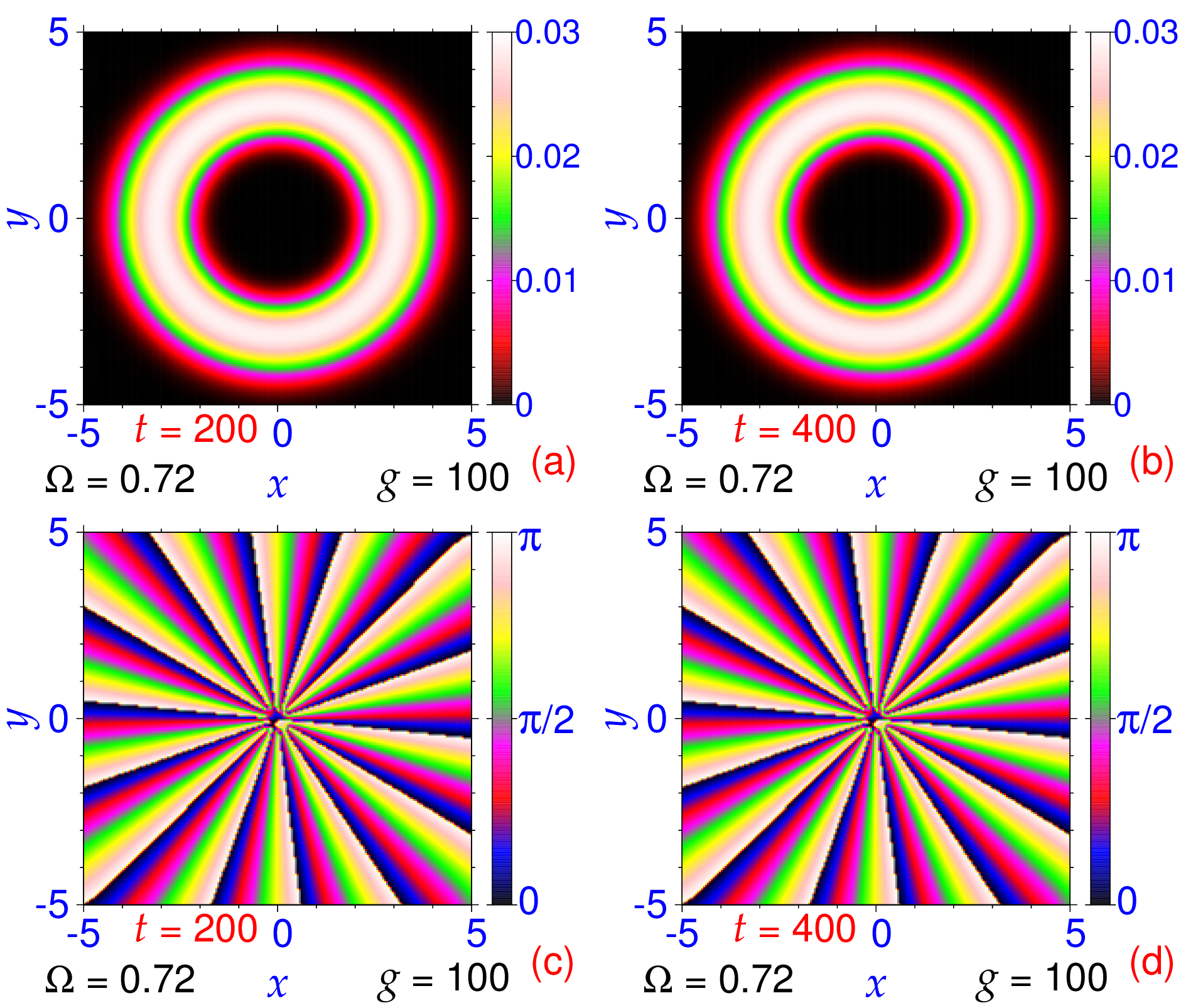}

\caption{  Dynamical evolution of  a giant vortex  of {the} rotating BEC displayed in figure \ref{fig3}(e), during  real-time propagation for 400 units of time
using the corresponding imaginary-time wave function as input, at times (a)
t = 200, (b) t = 400 for density. The corresponding phases are displayed in (c) and (d). During real-time propagation
the angular frequency of rotation $\Omega$ was changed at $t = 0$ from the imaginary-time value of $\Omega = 0.7$  to 0.72. 
}
\label{fig8}
\end{center}

\end{figure}

{  A giant vortex in an asymptotically harmonic trap is typically unstable \cite{uns}.} 
The dynamical stability of {  the present  quasi-2D  giant vortex}  is tested next. 
For this purpose we subject the giant-vortex  state  of the rotating BEC to real-time evolution during a large interval of
time using the initial stationary state as obtained by imaginary-time propagation, after slightly changing the angular frequency of rotation $\Omega$ at $t = 0$. The
giant vortex  will be destroyed after some time, if the underlying BEC wave function were dynamically
unstable. For this purpose, we consider a real-time propagation of the giant vortex  exhibited in
figure \ref{fig3}(e) for $g=100$ after changing $\Omega$ from 0.7 to 0.72 at $t = 0$. The subsequent evolution of
 the giant vortex  is displayed in figure \ref{fig8}  at  (a)  $t = 200$, (b) $t = 400$
for density and  {at} (c)  $t = 200$, (d) $t = 400$ for phase.
 The robust nature of the snapshots of the giant vortex  during
real-time evolution upon a small perturbation, as exhibited in figure \ref{fig8}, demonstrates
the dynamical stability of the giant vortex  in the quasi-2D rotating condensate.

\section{Summary and Discussion}

We have demonstrated the formation of a stable giant vortex in a controlled fashion, 
for the atomic interaction strength $g$  below a critical value,  in a rotating BEC 
trapped by a Mexican hat potential, which is a harmonic trap modulated by a small Gaussian hill at the center. The atomic interaction strength and the rotational frequency $\Omega$ can be kept very small which will make a  modeling of this phenomenology  extremely accurate and reliable and hence such a giant vortex can be used in high precision studies.  For   atomic interaction strengths above the critical value, a giant vortex can be generated provided that the  height of the central Gaussian hill is increased. Previous suggestions \cite{cornell,engels,aftalion,simula} for the generation of a giant vortex employed a  large value of $g$ and $\Omega$ or employed multi-component BECs \cite{multi} and hence the generated giant vortex could not be controlled like the present one and 
might  not be appropriate for high precision studies. The dynamical stability of the present giant vortex was established by real-time propagation. With present experimental know-how these giant vortices can  be prepared and studied in a laboratory.

\section*{Acknowledgements}
\noindent

SKA thanks the Funda\c c\~ao de Amparo \`a Pesquisa do
Estado de S\~ao Paulo (Brazil) (Project: 
2016/01343-7) and the Conselho Nacional de Desenvolvimento Cient\'ifico e Tecnol\'ogico (Brazil) (Project:
303280/2014-0) for partial support.


\section*{References}

\begin{thebibliography}{99}

\bibitem{becexpt}   
Anderson M H, Ensher J R, Matthews M R, Wieman C E and Cornell E A 1995 {\it Science} {\bf 269}
198

 Davis K B, Mewes M O, Andrews M R, van Druten N J, Durfee D S, Kurn D M and Ketterle
W 1995 {\it Phys. Rev. Lett.} {\bf 75} 3969 

{

\bibitem{yuka}
Courteille P W,  Bagnato V S and  Yukalov V I 2001 {\it Laser Phys.} {\bf 11}  659

 Yukalov V I 2004 {\it Laser Phys. Lett.}  {\bf 1}  435

Dalfovo F,  Giorgini S,  Pitaevskii L P and  Stringari S 1999
{\it Rev. Mod. Phys.} {\bf 71} 463 

}

\bibitem{vors}Madison K W, Chevy F, Wohlleben W and Dalibard J 2000 {\it Phys. Rev. Lett.} {\bf 84} 806

Matthews M R, Anderson B P, Haljan P C, Hall D S, Holland M J, Williams J E, Wieman
C E and Cornell E A 1999 {\it Phys. Rev. Lett.} {\bf 83} 3358

\bibitem{abri}Abrikosov A A 1957 {\it Zh. Eksp. Teor. Fiz.} {\bf 32} 1442 [Eng. Transla. 1957 {\it Sov. Phys.-JETP} {\bf 5} 1174]


\bibitem{fetter}Fetter A L 2009 {\it Rev. Mod. Phys.} {\bf 81} 647

\bibitem{4he} Yarmchuk E J and  Packard R E 1982  {\it J. Low Temp.
Phys.} {\bf 46}  479




\bibitem{vorl}Abo-Shaeer J R, Raman C, Vogels J M and Ketterle W 2001 Science 292 476

Abo-Shaeer J R, Raman C and Ketterle W 2002 {\it Phys. Rev. Lett.} {\bf 88} 070409



Haljan P C, Anderson B P, Coddington I and Cornell E A 2001 {\it Phys. Rev. Lett.} {\bf 86} 2922


 
 \bibitem{GP}
Gross E P 1961 {\it Nuovo Cimento} {\bf 20} 454

Pitaevskii L P 1961 {\it Zh. Eksp. Teor. Fiz.} {\bf 40} 646 [Eng. Transla. 1961 {\it Sov. Phys. JETP} {\bf 13} 451]

\bibitem{sci}  Adhikari S K  and  Salasnich L 2018 {\it Scientific Rep.} {\bf 8} 8825 



\bibitem{Schweikhard}Schweikhard V, Coddington I, Engels P, Tung S and Cornell E A 2004 {\it Phys. Rev. Lett.} {\bf 93}
210403
  

\bibitem{kumar}Kishor Kumar R, Tomio L, Malomed B A and Gammal A 2017 {\it Phys. Rev.} A {\bf 96} 063624

 Ghazanfari N,  Kele\c{s} A and    Oktel M  \"{O} 2014
{\it Phys. Rev.} A {\bf  89} 025601 


\bibitem{honey} Kasamatsu K and  Sakashita K 2018
{\it Phys. Rev.} A {\bf 97} 053622 





\bibitem{coreless} Leanhardt A E,  Shin Y,  Kielpinski D,  Pritchard D E and  Ketterle W 2003
{\it Phys. Rev. Lett.} {\bf   90}  140403 




\bibitem{frac1}  Su S-W,
 Hsueh C-H,
 Liu I-K,
 Horng T-L,
 Tsai Y-C,
 Gou S-C,
and  Liu W M 2011 {\it Phys. Rev.}  A {\bf 84}  023601 

 Cipriani M and  Nitta M  2013 
{\it Phys. Rev. Lett.} {\bf 111} 170401 



 \bibitem{cns} Adhikari S K 2019 {\it Commun. Nonlinear Sci. Numer. Simulat.} {\bf 71}  212 


\bibitem{thlatsep}  Mason  P and  Aftalion A 2011 
{\it Phys. Rev.} A {\bf  84} 033611 


\bibitem{cornell} Engels P,  Coddington I,  Haljan P C,  Schweikhard V and  Cornell E A 2003
{\it Phys. Rev. Lett.} {\bf 90} 170405 

\bibitem{simula} Simula T P,  Penckwitt A A and  Ballagh R J 2004
{\it Phys. Rev. Lett.} {\bf 92} 060401 



\bibitem{engels} Fetter A L,  Jackson B and  Stringari S 2005 {\it Phys. Rev.} A {\bf 71} 013605 

{ 
Kasamatsu K,  Tsubota M and  Ueda M 2002
{\it Phys. Rev.} A {\bf 66} 053606 



\bibitem{aftalion}Aftalion A and Danaila I 2003 {\it Phys. Rev.} A {\bf 69} 033608

\bibitem{kasamatsu} Cozzini M,  Jackson B and  Stringari  S 2006 \PR A {\bf 73}  013603

}







\bibitem{multi}
 Qin J,  Dong G and  Malomed B A 2016
{\it Phys. Rev.} A {\bf 94} 053611  

{ 
 Li Y,  Chen Z,  Luo Z,  Huang C,  Tan H,  Pang W and  Malomed B A 2018 \PR A {\bf 98} 063602 
}

Kuopanportti P,  Orlova N V and  Milo\v sevi\'c M V  2015
{\it Phys. Rev.} A {\bf 91} 043605 

Dong B,  Sun Q,  Liu W-M,  Ji A-C,  Zhang X-F and  Zhang S-G 2017
{\it Phys. Rev.} A {\bf 96} 013619 

 Xu X-Q and  Han J  H 2011
{\it Phys. Rev. Lett.} {\bf 107} 200401


\bibitem{detuned} Grimm R, Weidem\"uller M, Ovchinnikov Y B 2000 
{\it Adv. At. Mol.  Opt. Phys.} {\bf 42} 95

 He X, Yu S, Xu P,  Wang J and 
  Zhan M 2012 {\it Opt. Express} {\bf 20} 3711





\bibitem{onsager}Onsager L 1949   {\it Nuovo Cimento.} {\bf 6}  249, supp 2



\bibitem{feynman}  Feynman R P 1955  {\it Prog. Low Temp. Phys.} {\bf 1}  17


 \bibitem{Sonin}Sonin E B 2016 {\it Dynamics of Quantised Vortices in Superfluids} (Cambridge, Cambridge
University Press)


\bibitem{phase}   London F 1938  
{\it Nature}  {\bf 141} 643 

 
\bibitem{imag} Muruganandam P and Adhikari S K 2009 {\it Comput. Phys. Commun.} {\bf 180} 1888

 Vudragovi\'c D, Vidanovi\'c I, Bala\v z A, Muruganandam P and Adhikari S K 2012 {\it Comput. Phys.
Commun.} {\bf 183} 2021


Young-S. L E, Muruganandam P, Adhikari S K, Lon\v car
  V, Vudragovi\'c
 D and Bala\v z A 2017
{\it Comput. Phys. Commun.} {\bf
220}
 503


Young-S. L E, Vudragovi\'c D, Muruganandam P, Adhikari S K and Bala\v z A 2016 Comput.
Phys. Commun. 204 209

\bibitem{ll1960}Landau L D and Lifshitz E M 1960 {\it Mechanics} (Oxford, Pergamon Press) section 39

 

\bibitem{luca} Salasnich L, Parola A and Reatto L 2002 {\it Phys. Rev.} A {\bf 65} 043614



 
 
 

  
 

   
\bibitem{CPC} 
Lon\v car V, Bala\v z A, Bogojevi\'c A, Skrbi\'c S, Muruganandam P and Adhikari S K 2016
{\it
Comput.
Phys. Commun.} 
200 406


Satari´c B, Slavni\'c V, Beli\'c A, Bala\v z A, Muruganandam P and Adhikari S K 2016 Comput.
Phys. Commun. 200 411
 
\bibitem{cpckk}Kishor Kumar R, Lon\v car V, Muruganandam P, Adhikari S K and Bala\v z A 2019 {\it Comput. Phys.
Commun.} {\bf 240} 74
  

\bibitem{fesh}
 Inouye S,  Andrews M R,  Stenger J,  Miesner H-J,  Stamper-Kurn D M and  
Ketterle W 1998  {\it Nature}  {\bf   392} 151  

 
{ 
\bibitem{uns}
 Kuopanportti  P and  M\"ott\"onen M  2010
\PR A {\bf 81} 033627

 Kuopanportti  P and  M\"ott\"onen M 2010 {\it J. Low Temp. Phys.} {\bf 161} 561
}




\end{thebibliography}
\end{document}